\documentclass[3p,times]{elsarticle}

\usepackage{ecrc}

\usepackage{epstopdf}

\volume{00}

\firstpage{1}

\journalname{Nuclear Physics A}

\runauth{J. Noronha-Hostler et al.}

\jid{nupha}

\jnltitlelogo{Nuclear Physics A}

\usepackage{graphicx}
\usepackage{amsmath,amssymb}
\begin{document}

\begin{frontmatter}

%% Title, authors and addresses

%% use the tnoteref command within \title for footnotes;
%% use the tnotetext command for the associated footnote;
%% use the fnref command within \author or \address for footnotes;
%% use the fntext command for the associated footnote;
%% use the corref command within \author for corresponding author footnotes;
%% use the cortext command for the associated footnote;
%% use the ead command for the email address,
%% and the form \ead[url] for the home page:
%%
%% \title{Title\tnoteref{label1}}
%% \tnotetext[label1]{}
%% \author{Name\corref{cor1}\fnref{label2}}
%% \ead{email address}
%% \ead[url]{home page}
%% \fntext[label2]{}
%% \cortext[cor1]{}
%% \address{Address\fnref{label3}}
%% \fntext[label3]{}

\title{Understanding the $p/\pi$ ratio at LHC due to QCD mass spectrum}

%% Single author (and collaboration) - please insert
\author[a1]{Jacquelyn Noronha-Hostler}
\author[a2]{Carsten Greiner}
\address[a1]{Instituto de F\'{i}sica, Universidade de S\~{a}o Paulo, C.P.
66318, 05315-970 S\~{a}o Paulo, SP, Brazil}
\address[a2]{Institut f\"ur Theoretische Physik, Goethe Universit\"at, Frankfurt, Germany}

%% For multiple authors, replace the above by:

%\author[label1]{Author1}
%\author[label2]{Author2}

%\address[label1]{Address 1}
%\address[label2]{Address 2}

\begin{abstract}
Thermal fits have consistently reproduced the experimental particles yields of heavy ion collisions, however, the proton to pion ratio from ALICE Pb+Pb $\sqrt{s_{NN}}=2.76$ TeV is over-predicted by thermal models- known at the $p/\pi$ puzzle.  Here we test the relevance of the extended mass spectrum, i.e., include Hagedorn states (resonances that follow an exponential mass spectrum and have very short life times) on the $p/\pi$ puzzle.  We find that the extended mass spectrum is able to reproduce particle ratios at both RHIC and the LHC as well as being able to match the lower $p/\pi$ ratio at the LHC through dynamical chemical equilibration.  
\end{abstract}

\begin{keyword}
%% keywords here, in the form: keyword \sep keyword
hadron resonance gas \sep relativistic heavy-ion collisions \sep extended mass spectrum \sep Hagedorn states \sep dynamical hadronic interations \sep thermal fits
%% MSC codes here, in the form: \MSC code \sep code
%% or \MSC[2008] code \sep code (2000 is the default)

\end{keyword}

\end{frontmatter}

%%
%% Start line numbering here if you want
%%
% \linenumbers

%% main text

\section{Introduction}
\label{intro}

Final state particle ratios and yields have been matched using thermal fit models \cite{thermalmodels,NoronhaHostler:2009tz} in order to determine the chemical freeze-out temperature and baryonic chemical potential to aid in describing the nuclear phase diagram \cite{freezeoutline}.  Heavy ion experiments such as RHIC and SPS have found very precise matching to thermal models.  Using previous results, predictions were then made for the LHC \cite{LHC}.  However, recent results at ALICE Pb+Pb $\sqrt{s_{NN}}=2.76$ TeV for the LHC have proven to be difficult to fit by thermal models and the thermal models consistently overpredict the proton to pion ratio, $p/\pi$ - known as the $p/\pi$ puzzle \cite{Abelev:2012wca}. 

Various attempts have been made in understanding this puzzle  \cite{previous} and an overview of these attempts can be found in \cite{florisQM2014}. However, these attempts have not considered the effects of the extended mass spectrum (exponentially increasing mass spectrum of yet to be measured resonances with very short lifetimes) and/or multi-mesonic reactions. Evidence for the extended mass spectrum is found through the exponential behavior of the known hadrons \cite{Broniowski:2004yh}.   Previous work on the extended mass spectrum has found that they play a large role in the context of heavy ion collisions.  Most significantly, the extended mass spectrum decreases the shear viscosity to entropy density ratio in the hadron gas phase close to the AdS/CFT limit near to the critical temperature \cite{NoronhaHostler:2008ju,NoronhaHostler:2012ug,Pal:2010es}.  The extended mass spectrum is also able to extend the region of the thermodynamical quantities that we are able to match with lattice QCD \cite{Borsanyi:2010cj} from around $T=130-140$ MeV (for the standard hadron resonance gas) up to $T\approx 155$ MeV when the added resonances are present \cite{NoronhaHostler:2012ug,Majumder:2010ik}.  Elliptical flow is also suppressed in hybrid model calculations due to the extended mass spectrum when switching temperatures of $T_{SW}\approx155$ MeV or higher are used \cite{Noronha-Hostler:2013ria}.  Other effects of the extended mass spectrum have been found: an improvement of thermal fits \cite{NoronhaHostler:2009tz}, the phase change order \cite{phase}, and cumulants and correlations of charge fluctuations \cite{Bazavov:2014xya}.
 
Here we include the effects of the extended mass spectrum in order to explain the $p/\pi$ ratio at the LHC. Dynamical chemical reactions are used that are catalyzed by quickly decaying non-strange, mesonic Hagedorn states \cite{Hagedorn:1965st}.  We find that adding in the effects of the extended mass spectrum using out-of-chemical equilibrium dynamics can, indeed, explain the lower $p/\pi$ ratio at the LHC.

 \begin{figure}
\begin{center}
\includegraphics*[width=8.cm]{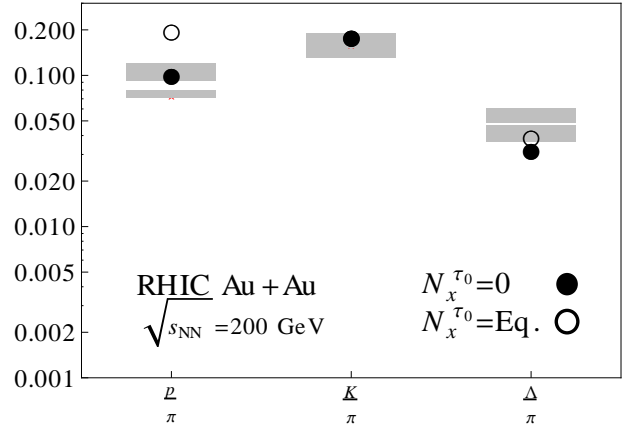}
\caption{
Comparison of the results of our extended mass spectrum model particle ratio calculations vs. experimental data points of both PHENIX and STAR at Au+Au RHIC $\sqrt{s_{NN}}=200$ GeV. The }
\label{fig:rhic}
\end{center}
\end{figure}

\section{Setup}
\label{setup}

The standard modeling of heavy-ion collisions has the following pattern: initial conditions until a time of $\tau_0\approx0.5-1$ fm, relativistic hydrodynamics is then initiated and allowed to expand and cool until the fluid cells reach a switching temperature of $T_{SW}\approx155$ MeV, once the fluid is converted into particles it is described by either a hadronic afterburner or hadronic transport model until chemical and kinetic freezeout is reached. For simplicity's sake, we begin a Bjorken expansion  with an accelerating radial flow that begins at $\tau_0$ using
\begin{equation}\label{eqn:bjorken}
V(\tau)=\pi\;\tau\left(r_{0}+v_{0}(\tau-\tau_{0})+\frac{1}{2}a_{0}(\tau-\tau_{0})^2 \right)^2
\end{equation}
to describe the relativistic hydrodynamical expansion.  Here we use the initial radius size of$r_0=7.1 \;fm$, and initial flow $v_{0}=0$ for both RHIC and LHC, whereas initial time is $\tau_{0}=0.6$ and $1.0$ fm and the acceleration is $a_{0}=0.03$ and $0.02 \;fm^{-1}$  for LHC and RHIC, respectively.  We ensure that causality is preserved and the final velocity is reasonable ($v_{final}\approx0.5-0.7$). We assume $T_{SW}=155$ MeV for both  LHC and RHIC, which corresponds to the temperature region where we expect the extended mass spectrum to be valid  \cite{NoronhaHostler:2012ug,Majumder:2010ik,Noronha-Hostler:2013ria}.  

After the switching temperature, we populate the hadrons using multihadronic decay reactions driven through Hagedorn states. To do so we must first establish the form of the mass spectrum, which is used to calculate the thermodynamics and chemical equilibrium values of the hadrons. 
In this proceedings we consider only the simplest form of the Hagedorn spectrum
\begin{equation}
\rho=A e^{m/T_{H}}\label{eqn:rho1}
\end{equation}
where $A=2.84 (1/GeV)$ and $T_H= 0.252$ GeV. However, in \cite{Noronha-Hostler:2014usa} we explore different descriptions of the extended mass spectrum, which also describe different types of decays.  The Hagedorn state decays that catalyze the other hadrons to quickly reach chemical equilibrium are
\begin{eqnarray}\label{eqn:decay}
n\pi&\leftrightarrow &HS\leftrightarrow n^{\prime}\pi+X\bar{X}.
\end{eqnarray} 
 where a Hagedorn state can either decay into multiple $n$ pions or $n^{\prime}$ pions plus an $X\bar{X}$ where $X\bar{X}=p\bar{p}$, $K\bar{K}$, or $\Lambda\bar{\Lambda}$. The exact rate equations used to describe these decays can be found in \cite{Noronha-Hostler:2014usa} with further details and tests of the robustness of our current assumptions.

\section{Results}
\label{results}

\begin{figure}
\begin{center}
\includegraphics*[width=8.cm]{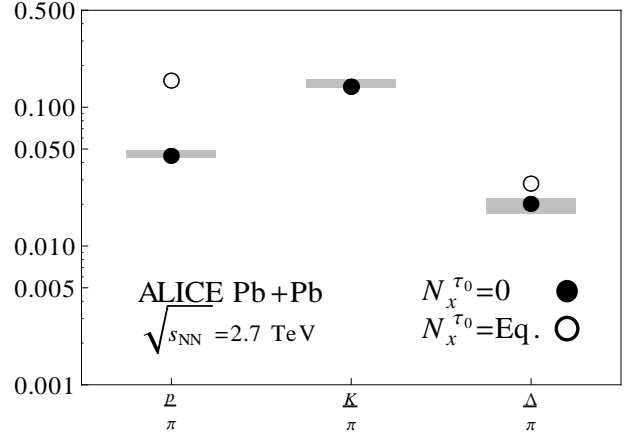}
\caption{
Comparison of the results of our extended mass spectrum model particle ratio calculations vs. experimental data points of ALICE at Pb+Pb LHC $\sqrt{s_{NN}}=2.76$ TeV.}
\label{fig:lhc}
\end{center}
\end{figure}

In Fig.\ \ref{fig:rhic} our results are shown compared to  Au+Au RHIC $\sqrt{s_{NN}}=200$ GeV data and in Fig.\ \ref{fig:lhc} our results are shown in comparison to Pb+Pb LHC $\sqrt{s_{NN}}=2.76$ TeV data.  The solid black dots represent the situation where there are no initial protons, kaons, and lambdas in our system (while the pions and Hagedorn states begin in chemical equilibrium) whereas the outlined circles represent the scenario when all hadrons begin in chemical equilibrium.  Follow these initial conditions the hadrons are then allowed to dynamically equilibrate over the expansion period. The LHC calculations end at $T_{end}=133$ MeV and the RHIC calculations end at $T_{end}=135$ MeV.  We caution against using $T_{end}$ MeV as a chemical equilibration temperature because it is highly dependent on the choice of parameters when describing the extended mass spectrum \cite{Noronha-Hostler:2014usa}. 

In a previous papers \cite{Greiner:2004vm,NoronhaHostler:2007jf,NoronhaHostler:2009cf} we explored the effect of the extended mass spectrum at RHIC and found that they were able to match experimental data points.  However, we have since updated our mass spectrum to fit to more recent lattice data \cite{Borsanyi:2010cj} and have also updated our modeling method to reflect the current modeling procedures of heavy ion collisions.  Thus, it is important to note that even after these changes the extended mass spectrum is able to match experimental particle ratios at RHIC (when the protons begin underpopulated).  This is an essential point because the same procedure that is used to match particle yields at RHIC using thermal fits was not able to adequately explain the lower $p/\pi$ ratio at LHC \cite{Abelev:2012wca,LHC}.

One can see in Fig.\ \ref{fig:lhc} that once again we are able to match the experimental data points at the LHC, which includes the lower $p/\pi$ ratio that has been unexplained by thermal fits \cite{Abelev:2012wca,LHC}.  However, this is only possible in the scenario  where there is an initial underpopulation of protons and lambdas.  When the protons  start in chemical equilibrium, then the proton to pion ratio is significantly overpopulated, which implies that both the extended mass spectrum combined with dynamical chemical equilibration is needed to explain the $p/\pi$ puzzle at the LHC.

\section{Conclusions}
 
Our results indicate that the inclusion of the  extended mass spectrum into dynamical hadron gas  interactions can then explain the suppressed $p/\pi$ ratio at the LHC.  Our current model is somewhat limited in that we can only consider non-strange, mesonic Hagedorn states.  We hope in the future to combine the transport model that includes the extended mass spectrum \cite{Beitel:2014kza} with a relativistic event-by-event hydrodynamical code \cite{vusphydro} to provide a more systematic check of the effect of the extended mass spectrum on the particles yields and collective flow.  We also plan on exploring the effect of the extended mass spectrum on multi-strange particles and additionally how strange and/or baryonic Hagedorn states would populate those states.

However, these results provide a strong indication that we could be seeing an effect from the extended mass spectrum at the LHC.  Already the extended mass spectrum plays a role in the transport coefficients, specifically the shear viscosity to entropy density ratio \cite{NoronhaHostler:2008ju, NoronhaHostler:2012ug}, and that additional strange baryons can affect cumulants and correlations of charge fluctuations \cite{Bazavov:2014xya}.  Thus, it is natural to question what other effects the extended mass spectrum will play on the signals of the Quark Gluon Plasma once they are integrated into hadronic after burners and transport methods. 

J.~Noronha-Hostler acknowledges
Funda\c{c}\~{a}o de Amparo \`{a} Pesquisa do Estado de S\~{a}o Paulo
(FAPESP)  for financial support. This work was supported by the Bundesministerium
f¨ur Bildung und Forschung (BMBF), the HGS-HIRe and
the Helmholtz International Center for FAIR within the
framework of the LOEWE program launched by the
State of Hesse.

%% The Appendices part is started with the command \appendix;
%% appendix sections are then done as normal sections
%% \appendix

%% \section{}
%% \label{}

%% References
%%
%% Following citation commands can be used in the body text:
%% Usage of \cite is as follows:
%%   \cite{key}         ==>>  [#]
%%   \cite[chap. 2]{key} ==>> [#, chap. 2]
%%

%% References with BibTeX database:

%\bibliographystyle{elsarticle-num}
%\bibliography{<your-bib-database>}

\begin{thebibliography}{99}

%\cite{Noronha-Hostler:2014usa}


\bibitem{thermalmodels}
  %%CITATION = ARXIV:0903.4379;%%
  P.~Braun-Munzinger, K.~Redlich and J.~Stachel,
  %``Particle production in heavy ion collisions,''
  arXiv:nucl-th/0304013;
  P.~Braun-Munzinger, J.~Stachel, J.~P.~Wessels and N.~Xu,
  %``Thermal equilibration and expansion in nucleus-nucleus collisions at the
  %AGS,''
  Phys.\ Lett.\  B {\bf 344}, 43 (1995);Phys.\ Lett.\  B {\bf 365}, 1 (1996);
  J.~Cleymans, D.~Elliott, A.~Keranen and E.~Suhonen,
  %``Thermal model analysis of particle ratios at GSI Ni Ni experiments  using
  %exact strangeness conservation,''
  Phys.\ Rev.\  C {\bf 57}, 3319 (1998);
  J.~Cleymans, H.~Oeschler and K.~Redlich,
  %``Influence Of Impact Parameter On Thermal Description Of Relativistic Heavy
  %Ion Collisions At (1-2) A-Gev,''
  Phys.\ Rev.\  C {\bf 59}, 1663 (1999);
  R.~Averbeck, R.~Holzmann, V.~Metag and R.~S.~Simon,
  %``Neutral pions and eta mesons as probes of the hadronic fireball in  nucleus
  %nucleus collisions around 1-A-GeV,''
  Phys.\ Rev.\  C {\bf 67}, 024903 (2003);
  P.~Braun-Munzinger, I.~Heppe and J.~Stachel,
  %``Chemical equilibration in Pb + Pb collisions at the SPS,''
  Phys.\ Lett.\  B {\bf 465}, 15 (1999).
  J.~Cleymans, H.~Satz, E.~Suhonen and D.~W.~von Oertzen,
  %``STRANGENESS PRODUCTION IN HEAVY ION COLLISIONS AT FINITE BARYON NUMBER
  %DENSITY,''
  Phys.\ Lett.\  B {\bf 242}, 111 (1990);
  J.~Cleymans and H.~Satz,
  %``Thermal hadron production in high-energy heavy ion collisions,''
  Z.\ Phys.\  C {\bf 57}, 135 (1993);
  F.~Becattini, M.~Gazdzicki and J.~Sollfrank,
  %``On chemical equilibrium in nuclear collisions,''
  Eur.\ Phys.\ J.\  C {\bf 5}, 143 (1998);
  F.~Becattini, J.~Cleymans, A.~Keranen, E.~Suhonen and K.~Redlich,
  %``Features of particle multiplicities and strangeness production in  central
  %heavy ion collisions between 1.7-A-GeV/c and 158-A-GeV/c,''
  Phys.\ Rev.\  C {\bf 64}, 024901 (2001); G.~Torrieri and J.~Rafelski,
  %``Strange hadron resonances as a signature of freeze-out dynamics,''
  Phys.\ Lett.\  B {\bf 509}, 239 (2001); G.~Torrieri, S.~Steinke, W.~Broniowski, W.~Florkowski, J.~Letessier and J.~Rafelski,
  %``SHARE: Statistical hadronization with resonances,''
  Comput.\ Phys.\ Commun.\  {\bf 167}, 229 (2005);
  S.~Wheaton and J.~Cleymans,
  %``THERMUS: A thermal model package for ROOT,''
  Comput.\ Phys.\ Commun.\  {\bf 180}, 84 (2009);
A.~Kisiel, T.~Taluc, W.~Broniowski and W.~Florkowski,
  %``THERMINATOR: Thermal heavy-ion generator,''
  Comput.\ Phys.\ Commun.\  {\bf 174}, 669 (2006) .
  C.~Spieles, H.~Stoecker and C.~Greiner,
  %``Hadron production in relativistic nuclear collisions: Thermal hadron
  %source or hadronizing quark-gluon plasma?,''
  Eur.\ Phys.\ J.\  C {\bf 2}, 351 (1998);
  B.~Schenke and C.~Greiner,
  %``Statistical description with anisotropic momentum distributions for  hadron
  %production in nucleus nucleus collisions,''
  J.\ Phys.\ G {\bf 30}, 597 (2004).
P.~Braun-Munzinger, D.~Magestro, K.~Redlich and J.~Stachel,
  %``Hadron production in Au Au collisions at RHIC,''
  Phys.\ Lett.\  B {\bf 518}, 41 (2001);
  W.~Florkowski, W.~Broniowski and M.~Michalec,
  %``Thermal analysis of particle ratios and p(T) spectra at RHIC,''
  Acta Phys.\ Polon.\  B {\bf 33}, 761 (2002);
  W.~Broniowski and W.~Florkowski,
  %``Strange particle production at RHIC in a single-freeze-out model,''
  Phys.\ Rev.\  C {\bf 65}, 064905 (2002);
  M.~Kaneta and N.~Xu,
  %``Centrality dependence of chemical freeze-out in Au + Au collisions at
  %RHIC,''
  arXiv:nucl-th/0405068;
  J.~Adams {\it et al.}  [STAR Collaboration],
  %``Experimental and theoretical challenges in the search for the quark  gluon
  %plasma: The STAR collaboration's critical assessment of the  evidence from
  %RHIC collisions,''
  Nucl.\ Phys.\  A {\bf 757}, 102 (2005).
A.~Andronic, P.~Braun-Munzinger and J.~Stachel,
  %``Hadron production in central nucleus nucleus collisions at chemical
  %freeze-out,''
  Nucl.\ Phys.\  A {\bf 772}, 167 (2006).
  %%CITATION = NUPHA,A772,167;%%
%\cite{Manninen:2008mg}
J.~Manninen and F.~Becattini,
  %``Chemical freeze-out in ultra-relativistic heavy ion collisions at
  %sqrt(s)_NN = 130 and 200 GeV,''
  Phys.\ Rev.\  C {\bf 78}, 054901 (2008).
  %%CITATION = PHRVA,C78,054901;%%
%\cite{NoronhaHostler:2009tz}
\bibitem{NoronhaHostler:2009tz} 
  J.~Noronha-Hostler, H.~Ahmad, J.~Noronha and C.~Greiner,
  %``Particle Ratios as a Probe of the QCD Critical Temperature,''
  Phys.\ Rev.\ C {\bf 82}, 024913 (2010)
  [arXiv:0906.3960 [nucl-th]].
  %%CITATION = ARXIV:0906.3960;%%
  %19 citations counted in INSPIRE as of 13 Aug 2013
\bibitem{freezeoutline}
  P.~Braun-Munzinger and J.~Stachel,
  %``Dynamics of ultra-relativistic nuclear collisions with heavy beams: An
  %experimental overview,''
  Nucl.\ Phys.\  A {\bf 638}, 3 (1998);
  J.~Cleymans and K.~Redlich,
  %``Unified description of freeze-out parameters in relativistic heavy ion
  %collisions,''
  Phys.\ Rev.\ Lett.\  {\bf 81}, 5284 (1998);Phys.\ Rev.\  C {\bf 60}, 054908 (1999);
  J.~Cleymans,
  %``Rapidity and energy dependence of thermal parameters,''
  J.\ Phys.\ G {\bf 35}, 044017 (2008);
  J.~Cleymans, R.~Sahoo, D.~K.~Srivastava and S.~Wheaton,
  %``Saturation of Transverse Energy per Charged Hadron and Freeze-Out Criteria
  %in Heavy-Ion Collisions,''
  Eur.\ Phys.\ J.\ ST {\bf 155}, 13 (2008)
  J.~Cleymans, R.~Sahoo, D.~P.~Mahapatra, D.~K.~Srivastava and S.~Wheaton,
  %``Transverse Energy per Charged Particle and Freeze-Out Criteria in Heavy-Ion
  %Collisions,''
  Phys.\ Lett.\  B {\bf 660}, 172 (2008);
  J.~Cleymans, H.~Oeschler, K.~Redlich and S.~Wheaton,
  %``Status of chemical freeze-out,''
  J.\ Phys.\ G {\bf 32}, S165 (2006).

\bibitem{LHC}
  N.~Armesto {\it et al.},
  %``Heavy Ion Collisions at the LHC - Last Call for Predictions,''
  J.\ Phys.\ G {\bf 35}, 054001 (2008).
  %%CITATION = JPHGB,G35,054001;%%
 
%\cite{Abelev:2012wca}
\bibitem{Abelev:2012wca} 
  B.~Abelev {\it et al.}  [ALICE Collaboration],
  %``Pion, Kaon, and Proton Production in Central Pb--Pb Collisions at $\sqrt{s_{NN}} = 2.76$ TeV,''
  Phys.\ Rev.\ Lett.\  {\bf 109}, 252301 (2012)
  [arXiv:1208.1974 [hep-ex]].
  %%CITATION = ARXIV:1208.1974;%%
  %30 citations counted in INSPIRE as of 27 Jul 2013
%\bibitem{Abelev:2013xaa} 
%  B.~B.~Abelev {\it et al.}  [ALICE Collaboration],
%  %``$K^0_S$ and $\Lambda$ production in Pb-Pb collisions at $\sqrt{s_{NN}}$ = 2.76 TeV,''
%  Phys.\ Rev.\ Lett.\  {\bf 111}, 222301 (2013)
%  [arXiv:1307.5530 [nucl-ex]].
%  %%CITATION = ARXIV:1307.5530;%%
%  %16 citations counted in INSPIRE as of 05 Mar 2014
  


%\cite{Steinheimer:2012rd}
\bibitem{previous} 
  J.~Steinheimer, J.~Aichelin and M.~Bleicher,
  %``Non-thermal $p/\pi$ ratio at LHC as a consequence of hadronic final state interactions,''
  Phys.\ Rev.\ Lett.\  {\bf 110}, 042501 (2013)
 .
  %%CITATION = ARXIV:1203.5302;%%
  %24 citations counted in INSPIRE as of 06 Aug 2013;C.~Ratti, R.~Bellwied, M.~Cristoforetti and M.~Barbaro,
  %``Are there hadronic bound states above the QCD transition temperature?,''
  Phys.\ Rev.\ D {\bf 85}, 014004 (2012)
  .
  %%CITATION = ARXIV:1109.6243;%%
  %28 citations counted in INSPIRE as of 25 Feb 2014

\bibitem{florisQM2014}
  M.~Floris, Quark Matter 2014 proceedings.
%\cite{Broniowski:2004yh}
\bibitem{Broniowski:2004yh} 
  W.~Broniowski, W.~Florkowski and L.~Y.~.Glozman,
  %``Update of the Hagedorn mass spectrum,''
  Phys.\ Rev.\ D {\bf 70}, 117503 (2004)
  [hep-ph/0407290].
  %%CITATION = HEP-PH/0407290;%%
  %52 citations counted in INSPIRE as of 13 Aug 2013
  



%\cite{NoronhaHostler:2008ju}
\bibitem{NoronhaHostler:2008ju} 
  J.~Noronha-Hostler, J.~Noronha and C.~Greiner,
  %``Transport Coefficients of Hadronic Matter near T(c),''
  Phys.\ Rev.\ Lett.\  {\bf 103}, 172302 (2009)
  [arXiv:0811.1571 [nucl-th]].
  %%CITATION = ARXIV:0811.1571;%%
  %77 citations counted in INSPIRE as of 13 Aug 2013
%\cite{NoronhaHostler:2012ug}
\bibitem{NoronhaHostler:2012ug} 
  J.~Noronha-Hostler, J.~Noronha and C.~Greiner,
  %``Hadron Mass Spectrum and the Shear Viscosity to Entropy Density Ratio of Hot Hadronic Matter,''
  Phys.\ Rev.\ C {\bf 86}, 024913 (2012)
  [arXiv:1206.5138 [nucl-th]].
  %%CITATION = ARXIV:1206.5138;%%
  %5 citations counted in INSPIRE as of 27 Jul 2013
  
  \bibitem{Pal:2010es} 
  S.~Pal,
  %``Shear viscosity to entropy density ratio of a relativistic Hagedorn resonance gas,''
  Phys.\ Lett.\ B {\bf 684}, 211 (2010)
 .
  %%CITATION = ARXIV:1001.1585;%%
  %19 citations counted in INSPIRE as of 27 May 2014
%\cite{Borsanyi:2010cj}
\bibitem{Borsanyi:2010cj} 
  S.~Borsanyi, G.~Endrodi, Z.~Fodor, A.~Jakovac, S.~D.~Katz, S.~Krieg, C.~Ratti and K.~K.~Szabo,
  %``The QCD equation of state with dynamical quarks,''
  JHEP {\bf 1011}, 077 (2010)
  [arXiv:1007.2580 [hep-lat]].
  %%CITATION = ARXIV:1007.2580;%%
  %276 citations counted in INSPIRE as of 27 Jul 2013
  %\cite{Majumder:2010ik}
\bibitem{Majumder:2010ik} 
  A.~Majumder and B.~Muller,
  %``Hadron Mass Spectrum from Lattice QCD,''
  Phys.\ Rev.\ Lett.\  {\bf 105}, 252002 (2010)
  [arXiv:1008.1747 [hep-ph]].
  %%CITATION = ARXIV:1008.1747;%%
  %14 citations counted in INSPIRE as of 28 Jul 2013
%\cite{Noronha-Hostler:2013ria}
\bibitem{Noronha-Hostler:2013ria} 
  J.~Noronha-Hostler, J.~Noronha, G.~S.~Denicol, R.~P.~G.~Andrade, F.~Grassi and C.~Greiner,
  %``Elliptic Flow Suppression due to Hadron Mass Spectrum,''
  Phys.\ Rev.\ C {\bf 89}, 054904 (2014).
  %%CITATION = ARXIV:1302.7038;%%
  %1 citations counted in INSPIRE as of 28 Jul 2013
  
  
  
\bibitem{phase} 
  L.~G.~Moretto, K.~A.~Bugaev, J.~B.~Elliott and L.~Phair,
  %``The Hagedorn thermostat,''
  Europhys.\ Lett.\  {\bf 76}, 402 (2006);
  V.~V.~Begun, M.~I.~Gorenstein and W.~Greiner,
  %``Crossover to Cluster Plasma in the Gas of Quark-Gluon Bags,''
  J.\ Phys.\ G {\bf 36}, 095005 (2009);
  I.~Zakout, C.~Greiner and J.~Schaffner-Bielich,
  %``The Order, shape and critical point for the quark-gluon plasma phase transition,''
  Nucl.\ Phys.\ A {\bf 781}, 150 (2007);
  I.~Zakout and C.~Greiner,
  %``The thermodynamics for hadronic gas of fireballs with internal color structures,''
  Phys.\ Rev.\ C {\bf 78}, 034916 (2008);
  L.~Ferroni and V.~Koch,
  %``Crossover transition in bag-like models,''
  Phys.\ Rev.\ C {\bf 79}, 034905 (2009);
  K.~A.~Bugaev, V.~K.~Petrov and G.~M.~Zinovjev,
  %``Quark Gluon Bags as Reggeons,''
  Phys.\ Rev.\ C {\bf 79}, 054913 (2009);
  A.~I.~Ivanytskyi, K.~A.~Bugaev, A.~S.~Sorin and G.~M.~Zinovjev,
  %``Critical exponents of the quark-gluon bags model with the critical endpoint,''
  Phys.\ Rev.\ E {\bf 86}, 061107 (2012)
  [arXiv:1211.3815 [nucl-th]].
  %%CITATION = ARXIV:1211.3815;%%
  %3 citations counted in INSPIRE as of 27 May 2014
  
  %\cite{Bazavov:2014xya}
\bibitem{Bazavov:2014xya} 
  A.~Bazavov, H.~-T.~Ding, P.~Hegde, O.~Kaczmarek, F.~Karsch, E.~Laermann, Y.~Maezawa and S.~Mukherjee {\it et al.},
  %``More strange hadrons from QCD thermodynamics and strangeness freeze-out in heavy ion collisions,''
  arXiv:1404.6511 [hep-lat].
  %%CITATION = ARXIV:1404.6511;%%
  
\bibitem{Hagedorn:1965st} 
  R.~Hagedorn,
  %``Statistical thermodynamics of strong interactions at high-energies,''
  Nuovo Cim.\ Suppl.\  {\bf 3}, 147 (1965);
R.~Hagedorn,
  %``Hadronic matter near the boiling point,''
  Nuovo Cim.\ A {\bf 56}, 1027 (1968).
\bibitem{Noronha-Hostler:2014usa} 
  J.~Noronha-Hostler and C.~Greiner,
  %``Suppression of the LHC $p/\pi$ ratio due to the QCD mass spectrum,''
  arXiv:1405.7298 [nucl-th].
  %%CITATION = ARXIV:1405.7298;%%
  
%%\cite{Frautschi:1971ij}
%\bibitem{Frautschi:1971ij} 
%  S.~C.~Frautschi,
%  %``Statistical bootstrap model of hadrons,''
%  Phys.\ Rev.\ D {\bf 3}, 2821 (1971).
%  %%CITATION = PHRVA,D3,2821;%%
%  %355 citations counted in INSPIRE as of 06 Nov 2013
%\bibitem{Greiner:1993jn} 
%  C.~Greiner, C.~Gong and B.~Muller,
%  %``Some remarks on pion condensation in relativistic heavy ion collisions,''
%  Phys.\ Lett.\ B {\bf 316}, 226 (1993).
  
  \bibitem{Greiner:2004vm}
  C.~Greiner {\it et al.}  
  %``Chemical equilibration due to heavy Hagedorn states,''
  J.\ Phys.\ G {\bf 31}, S725 (2005).
%\cite{NoronhaHostler:2007jf}
\bibitem{NoronhaHostler:2007jf} 
  J.~Noronha-Hostler, C.~Greiner and I.~A.~Shovkovy,
  %``Fast equilibration of hadrons in an expanding fireball,''
  Phys.\ Rev.\ Lett.\  {\bf 100}, 252301 (2008)
  [arXiv:0711.0930 [nucl-th]].
  %%CITATION = ARXIV:0711.0930;%%
  %31 citations counted in INSPIRE as of 27 Jul 2013
%\cite{NoronhaHostler:2009cf}
\bibitem{NoronhaHostler:2009cf} 
  J.~Noronha-Hostler, M.~Beitel, C.~Greiner and I.~Shovkovy,
  %``Dynamics of Chemical Equilibrium of Hadronic Matter Close to T(c),''
  Phys.\ Rev.\ C {\bf 81}, 054909 (2010)
  [arXiv:0909.2908 [nucl-th]].
  %%CITATION = ARXIV:0909.2908;%%
  %14 citations counted in INSPIRE as of 27 Jul 2013


%  
%  
%\bibitem{Abelev:2013vea} 
%  B.~Abelev {\it et al.}  [ALICE Collaboration],
%  %``Centrality dependence of $\pi$, K, p production in Pb-Pb collisions at $\sqrt{s_{NN}}$ = 2.76 TeV,''
%  Phys.\ Rev.\ C {\bf 88}, 044910 (2013)
%  
%  %\cite{Lizzi:1990na}
%\bibitem{Lizzi:1990na} 
%  F.~Lizzi and I.~Senda,
%  %``A Model of Interacting Strings and the Hagedorn Phase Transition,''
%  Phys.\ Lett.\ B {\bf 244}, 27 (1990).
%  %%CITATION = PHLTA,B244,27;%%
%  %28 citations counted in INSPIRE as of 27 May 2014
%  %\cite{Lizzi:1990if}
%\bibitem{Lizzi:1990if} 
%  F.~Lizzi and I.~Senda,
%  %``The Nucleation model of strings and the Hagedorn phase transition,''
%  Nucl.\ Phys.\ B {\bf 359}, 441 (1991).
%  %%CITATION = NUPHA,B359,441;%%
%  %12 citations counted in INSPIRE as of 27 May 2014
%  
%  
%\bibitem{Liu}
%  F.~M.~Liu, K.~Werner and J.~Aichelin,
%  %``Comparison of micro-canonical and canonical hadronization,''
%  Phys.\ Rev.\ C {\bf 68} (2003) 024905;
%  %%CITATION = HEP-PH 0304174;%%
%  F.~M.~Liu, et. al.,
%   %``A micro-canonical description of hadron production in proton proton
%  %collisions,''
%  J.\ Phys.\ G {\bf 30} (2004) S589;
%  Phys.\ Rev.\ C {\bf 69} (2004) 054002.



%\bibitem{Pal:2005rb} 
%  S.~Pal and P.~Danielewicz,
%  %``Hadron production from resonance decay in relativistic collisions,''
%  Phys.\ Lett.\ B {\bf 627}, 55 (2005)
%  [nucl-th/0505049].
%  %%CITATION = NUCL-TH/0505049;%%
%  %8 citations counted in INSPIRE as of 20 Feb 2014
%%\cite{Barannikova:2004rp}

  
   \bibitem{Beitel:2014kza} 
  M.~Beitel, K.~Gallmeister and C.~Greiner,
  %``Thermalization of Hadrons via Hagedorn States,''
  arXiv:1402.1458 [hep-ph].

%\cite{Noronha-Hostler:2013gga}
%\cite{Noronha-Hostler:2014dqa}
\bibitem{vusphydro}
J.~Noronha-Hostler, G.~S.~Denicol, J.~Noronha, R.~P.~G.~Andrade and F.~Grassi,
  %``Bulk Viscosity Effects in Event-by-Event Relativistic Hydrodynamics,''
  Phys.\ Rev.\ C {\bf 88}, 044916 (2013); J.~Noronha-Hostler, J.~Noronha and F.~Grassi,
  %``Bulk viscosity-driven suppression of shear viscosity effects on the flow harmonics at RHIC,''
  arXiv:1406.3333 [nucl-th].

\end{thebibliography}

%% Authors are advised to use a BibTeX database file for their reference list.
%% The provided style file elsarticle-num.bst formats references in the required Procedia style

%% For references without a BibTeX database:

\end{document}